**Pressure-dependent adhesion between solid-supported PC-lipid bilayers and vesicles under electric fields[†]**

Yu Zhang*, Di Jin, Jacob Klein*

Department of molecular chemistry and materials sciences, Weizmann Institute of Science, Rehovot, 7610001, Israel

* jacob.klein@weizmann.ac.il
* yu.zhang@weizmann.ac.il

**Abstract**

Fusion of lipid bilayers in membranes is important in processes from vesicle-cell interactions (as in drug delivery) to exosome-cell signaling, while transient transmembrane electric fields are known to occur spontaneously. Two contacting phosphatidylcholine (PC) lipid membranes are known to fuse into one under external electric fields, suggesting that the interaction between them is modified by the field as they approach, prior to the fusion event. Here we measure directly the adhesion energy between dimyristoylphosphatidylcholine (DMPC) and between distearoylphosphatidylcholine (DSPC) surface layers attached to solid substrates both without and with a transmembrane electric field. We find a marked pressure-dependent adhesion behavior in the electric field, which we attribute to fusion intermediates that are formed, shedding new light on membrane electro-fusion.

**Introduction**

Phosphatidylcholine (PC) lipids are one of the main components of the biological membrane systems, such as cell membranes, organelle membranes and exosomes, and play important roles in cell proliferative growth and programmed death [1, 2]. Liposomes or lipid nanoparticles, using PC lipid as carriers or protecting membranes, are widely used in drug delivery systems[3, 4], especially in vaccines [5]. PC-lipids are also found in healthy synovial fluids in the major joints (such as hips and knees) and on the surface of the articular cartilage layers coating the ends of joints, where they are the major lipid component (up to ca. 42% of total lipids [6, 7]). Indeed, the excellent lubrication of healthy articular cartilage, essential for the well-being of joints, has been attributed to PC lipid bilayers exposed at its outer surface [8]. In model surface forces balance (SFB) experiments, a friction coefficient down to $\mu \approx 10^{-4}$ is reported by Sorkin [9] and Goldberg [10] between PC-covered mica surfaces in physiological conditions. Such low friction is attributed, via the hydration lubrication mechanism [9], to the tenaciously-held hydration shells surrounding the PC-headgroups [11].

Tribological studies to date between lipid layers using the SFB [9, 10] were generally carried out under "zero" transverse electric fields, as the two opposing mica substrates bear identical surface charge density [12]. The presence of electric fields may considerably modify the bilayer structure (e.g. via electroporation or membrane fusion [13]) and thus the interaction between the bilayers. Moreover, recent work [14] suggests that significant transverse fields across cell membranes in vivo may arise spontaneously due to transient potential imbalance across the lipid bilayers. Any structural changes in the lipid membranes induced by such fields are expected also to influence the lateral frictional forces. Indeed in very recent studies, we found that transverse electric fields had a strong effect on the sliding friction between lipid bilayers, and this is attributed to field-induced topological changes such as electroporation [15-17].

It is also reasonable to expect that the adhesion energy, i.e. the energy required to separate two contacting lipid-bilayer bearing surfaces, might change in the presence of a transverse electric field, due to such changes in the bilayer structures, and that this would also depend on the applied pressure as this would further modulate these structures. In the present study, we explore the effects of electric field and normal pressure on adhesion between mica and gold substrates bearing dimyristoylphosphatidylcholine (DMPC) or distearoylphosphatidylcholine (DSPC) lipid layers. The DMPC lipids, close to their liquid-disordered phase at room temperature, are in the form of extended lipid bilayers (LBs) on the substrates (as their liposomes rupture when adsorbed on the surfaces), while the gel-phase DSPC vesicles retain their liposomic structure on the surfaces [9].

## 2. Materials and methods

2.1 Materials

Water used was purified (TOC<1 ppb and 18.2 MΩ•cm conductivity) using a Barnstead$^{TM}$ GenPure$^{TM}$ system (Thermal Scientific). Ruby muscovite mica (grade I) was purchased from S & J Trading, Inc. Gold and silver pellets (99.999%) were obtained from Kurt J. Lesker Inc. DMPC and DSPC are purchased from Lipoid (Germany) and used as received. Platinum wire (99.99%) for electrodes is obtained from Advent Ltd. All solvents were analytical grade (Merck, Sigma, and BioLab). EPON 1004 resin (Shell) is used to glue the gold and mica on fused hemi-cylindrical silica lenses.

2.2 Liposome extrusion and adsorption

The DMPC and DSPC liposomes are prepared using a standard extrusion method[18]. The lipid powder is vortex-dispersed in pure water at 5 mM concentration, then sonicated for 20 min at 10 °C above respective phase transition temperature $T_M$ ($T_M$ = 24 °C for DMPC and 55 °C for DSPC [19]) forming dispersions of multilamellar vesicles (MLVs). The MLVs are downsized by passing the dispersion through filters with designated pore radius for several cycles: 0.4 μm for 8 cycles, 0.1 μm for 10 cycles, and 0.05 μm for 12 cycles, to yield dispersions of small unilamellar vesicles (SUVs, or liposomes). The dispersion temperature is maintained during the downsizing process following which dispersions are kept at 4 °C for less than 12 hours before use. Gold-coated or bare mica surfaces (substrates) are incubated overnight in 0.5 mM liposome dispersion then washed before mounting into the SFB.

## 2.3 Surface force balance (SFB) experiments

Mica surfaces are prepared by cleaving mica sheets into ca. 2.5 μm thick single crystalline facets, as reported previously [20]. Then a ca. 60 nm thick reflective silver layer is evaporated on one side of the prepared facets. Molecularly smooth gold surfaces are prepared using the template-stripping method [21]. Briefly, a ca. 60 nm thick gold film is evaporated onto a cleaved single-crystal mica template (ca. 10 μm thick) at the rate of 0.1 Å/min (Odem evaporator), and annealed at 150 °C for 2 hours. This gold-coated mica sheet is then glued onto the SFB lens and the molecularly smooth gold surface in contact with the atomically-smooth mica is finally exposed by peeling the template off.

The SFB experiments are carried out as described in detail previously [22, 23], using a three-electrode modified SFB (Fig. 1). Briefly, the gold-exposing and the bare, back-silvered mica surfaces are mounted opposite to each other in a cross-cylinder configuration, with the gold surface (upper lens) mounted on a sectored piezoelectric tube and the mica substrate (lower lens), immersed in a water bath, on a normal spring set. The geometry is equivalent to a sphere (radius $R$ ca. 1 cm) interacting with a flat surface. A beam of white light is passed through the lenses, forming multiple-beam interference fringes (fringes of equal chromatic order, as in upper insert in Fig. 1) whose wavelength is measured to yield the gold-mica separation $D$ [24, 25]. The two surfaces are brought into contact using either a step-wise or a dynamic method and the normal forces $F_n$ between them are calculated from the bending of the normal leaf spring. Lateral motion is induced via opposing potentials on a sectored piezoelectric (PZ) tube, and the friction force $F_s$ measured as $F_s=K_s\Delta x$ where $K_s$ is the shear leaf spring constant and $\Delta x$ is the bending as measured by the air-gap-capacitor. The gold surface potential is controlled by a three-electrode configuration, consisting of gold as working electrode (W), and two platinum wires as counter (C) and quasi-reference electrodes (R). Normal force profiles between bare substrates are measured to calibrate the surface potential of gold using Poisson-Boltzmann equation [17]. The adhesion energy $W_{adh}$ is calculated using JKR contacting model from the "pull-off" force $F_{pulloff}$ upon separating two surfaces [26],

$$F_{pulloff} = (3/2)\pi W_{adh} R \qquad (1)$$

where $R$ is the mean radius of curvature of the two cylindrical surfaces, measured from the fringe shapes. In principle the adhesion energy as thus defined should be independent of the contact pressure (or contact area) between the surfaces, but for convenience we will use this definition (eq (1)) also for the present study where $W_{adh}$ (which we may also call the apparent adhesion energy) can depend on the applied pressure.

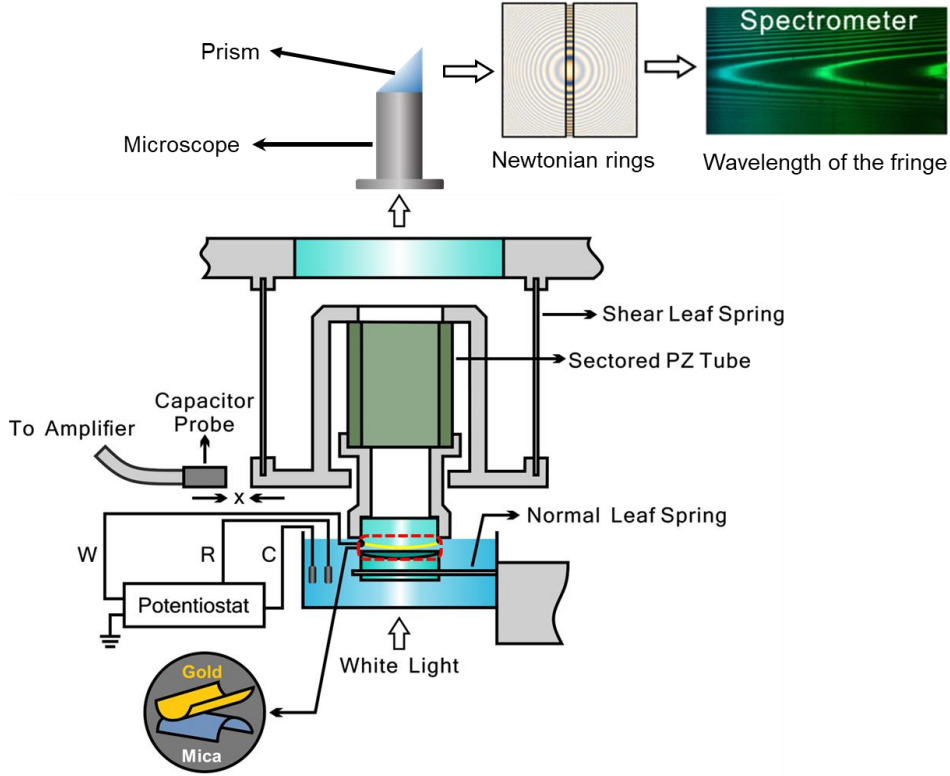

**Fig. 1:** Schematic of the three-electrode-modified surface force balance (SFB) showing the main parts: sectored piezo (PZ) tube, air-capacitor probe, normal and shear spring sets. The potential of the gold surface is controlled by a three-electrode configuration: the gold surface as working electrode (W), two platinum wires as quasi-reference electrode and counter electrode. Two fused silica lenses are mounted in cross-cylinder configuration (bottom left insert) that is equivalent to a sphere on a flat configuration. The lower circular inset indicates the geometry of the gold and mica substrates. Modified with permission from ref. [27]. Copyright 2021 AIP Publishing LLC.

## 3. Results

The approaching (increasing load) and receding (decreasing load) normal force $F_n(D)$ profiles (normalized as $F_n(D)/R$, in the Derjaguin approximation [28]) between PC-covered substrates are measured under an electric field induced by varying the potentials $\Psi_{app}$ applied to the gold surface; this is different to the actual potential $\Psi_{gold}$ at the surface which can be extracted from the force profiles as earlier discussed [23]. We consider two cases, as illustrated in Fig.2.

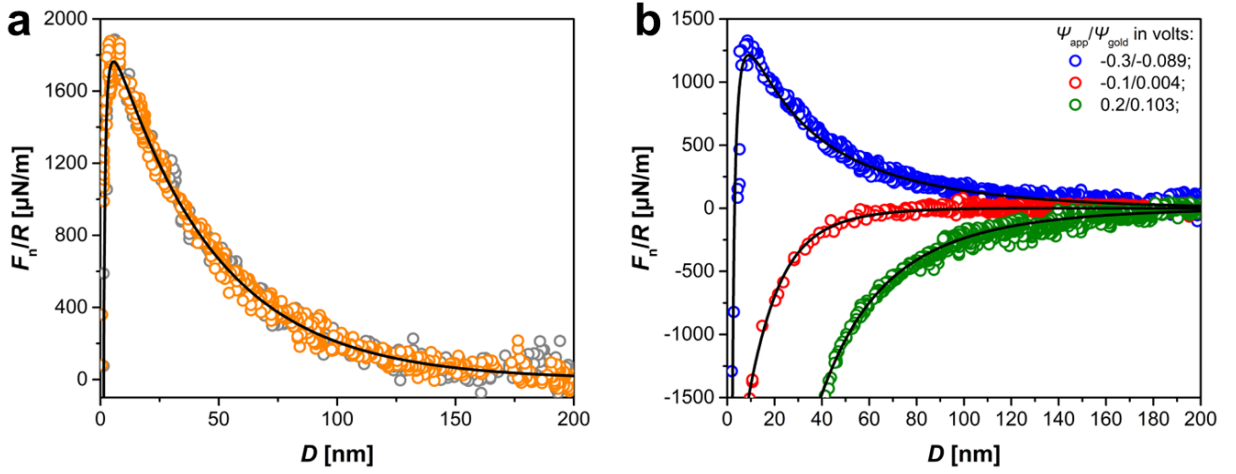

**Fig. 2:** Normal force $F_n(D)/R$ profiles between two mica surfaces (a), and between gold and mica surfaces (b), collected using the dynamic method [29]. These results are used to calculate the mica surface charge density and gold surface potentials through fitting the force profiles with Poisson-Boltzmann (Gouy-Chapman) model [28] (black solid lines in a and b). The boundary conditions are constant charge vs. constant charge (panel a) and constant charge vs. constant potential (panel b). The mica surface charge $\sigma_{mica}$=-5.21 mC/m$^2$ (corresponding to $\Psi_{mica,\infty}$=-0.095 V at infinite separation $D$ in $c_0$=7.8×10$^{-5}$ M 1:1 electrolyte solution) is extracted from panel a and used as constant in the fitting in panel b.

This shows the normal force profiles between two bare mica surfaces (Fig. 2a), from which a surface charge density on each surface can be extracted (the mica surfaces are at constant charge), corresponding to a mica surface potential $\Psi_{mica,\infty}$ = -0.095 V (at infinite separation $D$). The force profiles between bare mica and a bare gold surface at different applied potentials $\Psi_{app}$ are shown in fig. 2b: when $\Psi_{app}$=-0.3 V, the potential of the gold surface is $\Psi_{gold}$ = -0.089 V, very similar to that of the bare mica, implying a low electric field $E$ ($\approx$ 0) between them [23]. We designate that the low field case, and it is similar in terms of surface charges density and potential variation across the gap to the extensively-studied symmetric case of two interacting mica surfaces. When the applied potential is made more positive, $\Psi_{app}$= -0.1 V (red data) or 0.2 V (green data), the gold potential changes sign ($\Psi_{gold}$ = +0.004 V and +0.103 respectively) and a high field $E \approx 10^7 \sim 10^8$ V/m [17] results across the gold/mica gap; this is designated the high field case.

## 3.1 Forces between surfaces coated with DMPC-LBs

The approaching or loading (increasing load $F_n$) and receding or unloading (decreasing load) normal force profiles measured at the two different potentials $\Psi_{app}$=-0.3 V (low field), and $\Psi_{app}$=-0.1 V (high field) are shown in Fig. 3a and 3b respectively. The approaching force profiles in the low field case (Fig. 3a, full symbols) show a combination of long-range ($D>10$ nm) electrostatic interactions and short-range steric repulsion ($D \leq 10$ nm), qualitatively similarly to those in pure water between bare mica surfaces at different gold potentials (Fig. 2a). The long-range repulsions, fitted well by the Gouy-Chapman model (grey solid lines in 3a) [28], are of electrostatic origin arising from the osmotic pressure of trapped counterions. In contrast, the two oppositely charged surfaces (Fig. 3b) "jump-in" to contact at the Euler instability point when $\partial F_n/\partial D > K_n$ (black arrow in Fig. 3b). Short range interactions are dominated by steric repulsions between the DMPC-LBs, consistent with previous results[9]. Such repulsion results in a "hard-wall", at separation $D = D_{hw}$=8.6±0.8 nm (Fig. 3a) and $D_{hw}$=8.3±0.8 nm (Fig. 3b), indicating two confined lipid bilayers at contact [30].

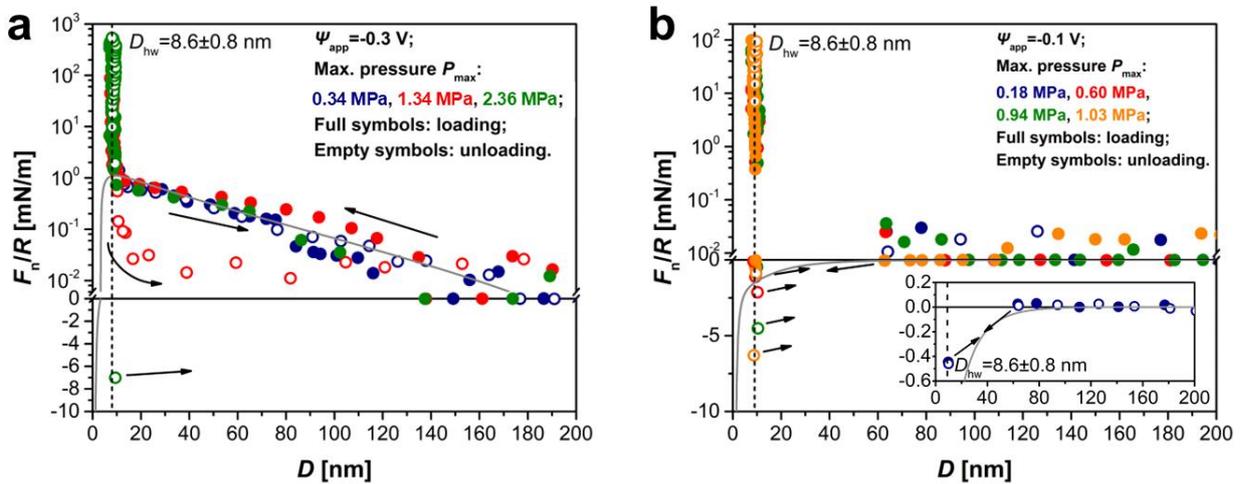

**Fig. 3:** Typical loading and unloading normal force profiles between DMPC-LB-covered substrates, plotted as $F_n(D)/R$ in the Derjaguin approximation: a) the low-field or 'repulsive' regime ($\Psi_{app}$=-0.3 V, $F_n/R > 0$ on approach at all separations); and b) the high-field or 'attractive' regime ($\Psi_{app}$=-0.1 V, $F_n/R < 0$ on approach). Full and empty symbols represent approaching and receding profiles, respectively. Different colors represent repeat measurements, in which the maximum applied pressure (same colour code as data symbols) is different. Grey solid curves in a is the best fit using a Gouy-Chapman model [28] with mica surface charge density $\sigma_{mica}$= -5.45

mC/m$^2$, gold surface potential are the same of cases in water, $\Psi_{gold}$= -0.089 V (a) and $\Psi_{gold}$= 0.004 V (b), bulk electrolyte concentration (assuming 1:1 salt) $c_0$= 7.0×10$^{-5}$ M, and gold-mica Hamaker constant $A_H$= -9×10$^{-20}$ J [22]. The insert in b shows magnification of blue data sets.

The receding force profiles (empty symbols in Fig. 3), show both a pressure and electric field-dependent behaviour. In both low and high field cases the surfaces are compressed to different maximal loads, corresponding to differing maximal contact pressures $P_{max}$ as indicated by the colour coding. In the low field regime (Fig. 3a) at low $P_{max}$, the approaching and receding profiles overlap with each other (blue data in Fig. 3a) with little hysteresis. On increasing $P_{max}$ (Fig.3a, red and green data) there is increasing hysteresis. The possible origins of this are considered in the Discussion section. For the high $E$-field case, Fig. 3b, we observe jumps into contact as the surfaces approach, as expected [23]. Here we see that as $P_{max}$ increases the pull-off force and thus the apparent adhesion energy increases. We note, importantly, that the data in Figs. 3a, b were all taken at a given contact point for each lipid type (to maintain the constant electric field at given potential), and that profiles were carried out successively going from lower to higher $P_{max}$, with ca. 15' wait time between successive loading cycles. The fact that the subsequent loading profiles reproduce themselves with little hysteresis (i.e. red data following the blue, and green following the red in Fig. 3a; and likewise in Fig. 3b with additionally orange following the green) suggests that the lipid layers largely recover in between loading processes from any compression induced changes at $P_{max}$, as further discussed below. Such measurements are carefully repeated also at different contact points, and similar results are acquired as summarized in Fig.4.

The adhesion energy (shown in Fig. 4), calculated from the pull-off forces according to eq (1), can be separated into two regimes. For the 'repulsive' low-field case (from data such as in Fig. 3a) there is no net adhesion at lower $P_{max}$ values (i.e. no pull-off force from adhesion; $W_{adh}$ = 0 and is independent of $P_{max}$; empty symbols in Fig. 4), while at highest $P_{max}$ there is substantial scatter in the pull-off force (and $W_{adh}$). In contrast, for the high field case (from data such as in Fig. 3b) the adhesion energy rises with the pressure $P_{max}$ (full symbols in Fig.4), showing pressure dependent behavior.

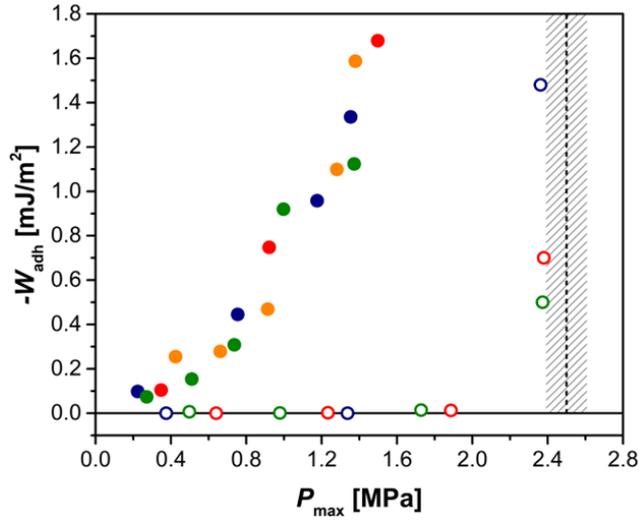

**Fig. 4:** Apparent adhesion energies evaluated from data such as in Fig. 3 and eq (1) between DMPC bilayer-covered gold and mica surfaces, in high electric field ($\Psi_{app}$= -0.1 V, full symbols), and low field ($\Psi_{app}$= -0.3 V, empty symbols). The data from repeat experiments at different contact points are shown in different colors. The vertical dashed lines represent the pressure for hemifusion at $\Psi_{app}$= -0.3 V, and the cross-hatched band shows its standard deviation.

3.2 Forces between surfaces coated with DSPC liposomes

The approaching force profiles of DSPC are monotonically repulsive (Fig.5, full symbols), irrespective of the electric field, similar to the previous results between two mica substrates [31], attributed to steric repulsion between the intact liposomes that is not influenced by the electric field [32, 33]. A "hard-wall" separation $D_{hw}$ = 20.4±0.8 nm and 20.0±0.8 nm, corresponding to 2 layers of flattened liposomes, i.e. 4 layers of LBs, are measured, indicating the integrity of DSPC-liposomes [34, 35]. The DSPC-LBs thickness is given by $D_{hw}/4$ = 5.1±0.2 nm and 5.0±0.2 nm (compared with the corresponding DMPC-LB thickness values of 4.3±0.2 nm and 4.2±0.2 nm respectively).

The receding force profiles of DSPC-liposomes (empty symbols in Fig. 5) show a similar hysteresis to those of DMPC-LBs. The loading/unloading hysteresis gradually appears as the maximum pressure $P_{max}$ increases, for DSPC-liposome both with and without electric field. This trend is observed for the low field, but is most clearly seen in the increasing pull-off forces when separating two surfaces in the high field case (Fig. 5b): the pull-off forces increase from zero (in the absence of an $E$-field, Fig. 5a) to finite increasing values as the pressure increases in the loading force profiles.

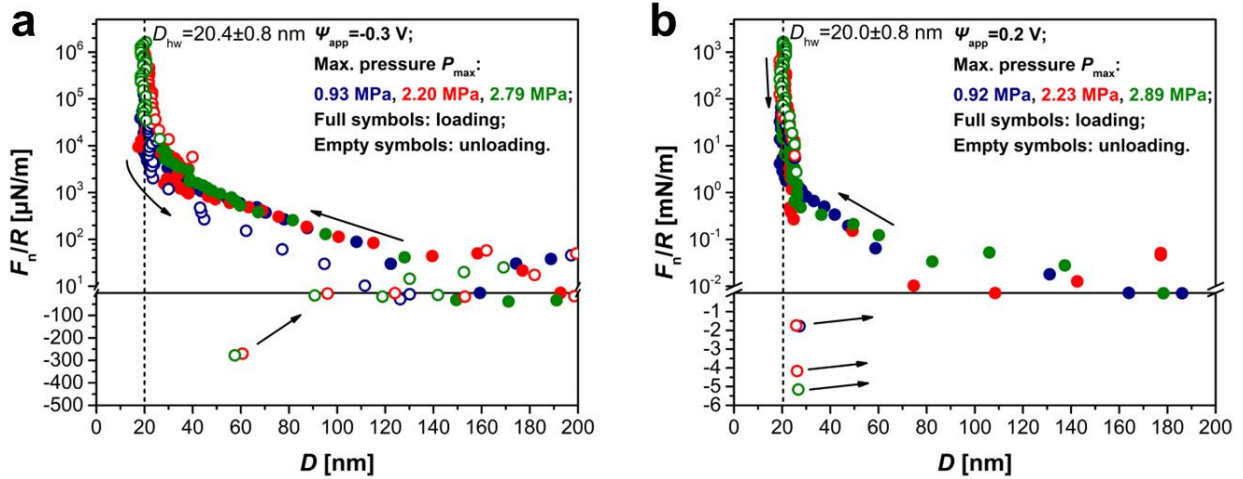

**Fig. 5:** Typical approaching (filled symbols) and receding (empty symbols) normal force profiles between DSPC-SUVs covered gold and mica surfaces, plotted as $F_n(D)/R$ in the Derjaguin approximation where $R$ is the mean curvature radius of the surfaces: a) the low-field or 'repulsive' regime ($\Psi_{app}$= -0.3 V); and b) the high-field or 'attractive' regime ($\Psi_{app}$= 0.2 V), measured at a given contact point to maintain the constant electric field at given potential. The maximum pressures at all loading profiles are represented by different colors, and its effect is summarized below.

The adhesion energies of DSPC-liposomes are plotted against the maximum pressures $P_{max}$, as shown in Fig. 6. Similar to the DMPC-LBs, two different adhesion types are clearly shown. At low electric field (empty symbols), the DSPC-liposome adhesion energy $W_{adh}$ = 0 and is independent of $P_{max}$, consistent with earlier studies [10, 31]. At high electric field (full symbols), a $P_{max}$-dependent apparent adhesion energy is clearly seen from the data, similar to those in DMPC-LBs (Fig. 3b and full symbols in Fig. 4).

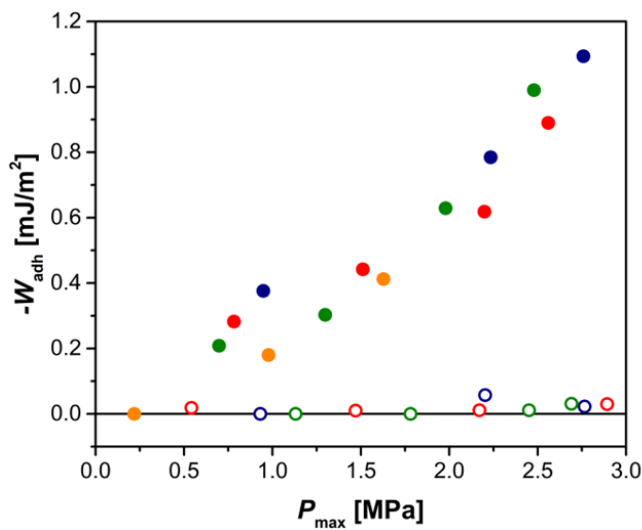

**Fig. 6:** Adhesion energies (evaluated from pull-off forces as in Fig. 5, via eq (1)) between DSPC-SUVs covered gold and mica surfaces, at attractive regime ($\Psi_{app}$= 0.2 V, full points), and repulsive regime ($\Psi_{app}$= -0.3 V, empty points). The data from repeat experiments at different contact points are shown in different colors. The results indicate that the adhesion energy increases with the maximum pressure when electropores appear under electric field.

## 4. Discussion

The main new finding in this study is the hysteretic pressure-dependent adhesion of PC-covered substrates, weakly in the case of low or zero field between the surfaces, and much more marked when a high field (ca. $10^7 – 10^8$ V/m) is applied between them [17, 36]. Pressure-dependent adhesion is common between *polymer*-coated surfaces (no electric field), due to the contacting of asperities from opposite surfaces as pressure increases [37-39]. A closer analogy for the case of zero transverse *E*-field with lipid layers is proposed by Sun et al. [40] and other groups [41-43] to explain adhesion between giant unilamellar vesicles and tethered LBs. They divide the adhesive behavior into a "low adhesion" regime, where no fusion intermediates forms, thus the adhesion is pressure-independent, and a "high adhesion" regime, where two LBs are connected by fusion intermediates (though they do not specify the nature of such 'fusion intermediates'). They attribute the stronger adhesion at higher normal pressures to the more favorable formation of fusion intermediates [40], thus additional energy is needed to separate two LBs, leading to pressure-dependent adhesion. These studies were all carried out in the absence of any applied electric fields, and are in line with our results as in Figs. 3a and 5a with no electric fields across the bilayers.

How can we explain the much stronger pressure dependency of the pull-off force (and thus of the apparent adhesion energy) seen in the high field case, shown in Figs. 3b and 5b, and summarized in Figs. 4 and 6, which is the most striking result of this study? Very recent work has shown that a transverse electric field of the same magnitude as in our experiments (in the high field case) induces electroporation and hydrophobic contacts between two PC LBs attached to gold and mica substrates [15-17]. Once these form due to such transmembrane $E$-fields, even at low applied pressures, they contribute to the adhesion as described above, and the stronger the pressure the more extensive the effect: this is because higher pressures lead to more dehydration of the LBs [44]. We thus attribute the strong adhesion hysteresis seen in the high field case in our system to this $E$-field-induced process initiating these topological changes. We emphasize that these structures form with little change in the bilayer thickness, which is indeed what we observe. At yet higher pressures we would expect hemifusion to occur for the near-liquid-phase DMPC-LBs, as earlier observed [45], with an abrupt expulsion of a bilayer; but this is not the case at the contact pressures of the present study, and indeed the lipid layers do not change their thickness significantly in our experiments.

We may now account also for the results showing adhesion hysteresis with increasing pressures also for the case of no electric fields, as observed in Figs. 3a and 5a and also in earlier studies [40-43]: the idea that hydrophobic contacts form between bilayers that are sufficiently compressed (and thus somewhat dehydrated relative to uncompressed LBs), can explain this relatively weak hysteresis. Once they form, they are likely to have an attractive contribution to the overall interaction as the surfaces separate on unloading, and this may account for the hysteretic behaviour and even the overall net adhesion as seen in Figs. 3a and 5a in the case of highest $P_{max}$ where their formation is likely to be most marked. Such hydrophobic contact formation and resulting hysteresis may also be related to the sharp increase in sliding friction observed by Sorkin et al. earlier [9] between mica surfaces with attached DMPC-LBs, at loads similar to or higher than those corresponding to the highest red data points in Fig. 3b (but not at the lower loads corresponding to the blue data, where the sliding friction remained very low, indicating unperturbed DMPC-LBs, consistent with the absence of such contacts or hysteretic effects).

## 5. Conclusions

The apparent adhesion energy (derived from the pull-off force via the contact mechanics relation) between surface-attached DMPC-lipid bilayers near their liquid phase, and between DSPC vesicles in their gel phase, are studied both with and without an externally-applied transverse electric field ($E$-field). We find that in the absence of an $E$-field the apparent interaction hysteresis on loading/unloading is pressure-independent at low contact pressures $P$ but increases with $P$ at higher pressures. However, when an $E$-field is applied, the apparent adhesion energy depends on the pressure already from low $P$ values. This behaviour is attributed to fusion intermediates such as electro-pores and hydrophobic contacts: these latter may form at high enough pressures even with no $E$-field, but in the presence of an $E$-field they may form spontaneously even in the absence of pressure. Once formed they proliferate at higher pressures due to progressive dehydration of the LBs, thereby providing a $P$-dependent attractive component to the inter-bilayer interactions. Our results thus have a clear connection to membrane interactions in biological systems.

**Acknowledgements**

We thank the European Research Council (Advanced Grant CartiLube 743016), the McCutchen Foundation, the Israel Science Foundation–National Natural Science Foundation of China joint research program (grant 3618/21), and the Israel Science Foundation (grant 1229/20) for financial support. This work was made possible partly through the historic generosity of the Perlman family.

**Conflict of interests**

The authors have no conflict of interests to declare.

**References**
1.  Ridgway, N.D., *The role of phosphatidylcholine and choline metabolites to cell proliferation and survival.* Crit Rev Biochem Mol Biol, 2013. **48**(1): p. 20-38.

2. Zhan, Q., et al., *Phosphatidylcholine-Engineered Exosomes for Enhanced Tumor Cell Uptake and Intracellular Antitumor Drug Delivery.* Macromol Biosci, 2021. **21**(8): p. e2100042.

3. Wang, W., et al., *Functional Choline Phosphate Lipids for Enhanced Drug Delivery in Cancer Therapy.* Chemistry of Materials, 2021. **33**(2): p. 774-781.

4. Gkionis, L., et al., *Microfluidic-assisted fabrication of phosphatidylcholine-based liposomes for controlled drug delivery of chemotherapeutics.* Int J Pharm, 2021. **604**: p. 120711.

5. Pardi, N., M.J. Hogan, and D. Weissman, *Recent advances in mRNA vaccine technology.* Curr Opin Immunol, 2020. **65**: p. 14-20.

6. Kosinska, M.K., et al., *A lipidomic study of phospholipid classes and species in human synovial fluid.* Arthritis Rheum, 2013. **65**(9): p. 2323-33.

7. Sarma, A.V., G.L. Powell, and M. LaBerge, *Phospholipid composition of articular cartilage boundary lubricant.* Journal of Orthopaedic Research, 2001. **19**(4): p. 671-676.

8. Jahn, S., J. Seror, and J. Klein, *Lubrication of Articular Cartilage.* Annual Review of Biomedical Engineering, 2016. **18**(1): p. 235-258.

9. Sorkin, R., et al., *Origins of extreme boundary lubrication by phosphatidylcholine liposomes.* Biomaterials, 2013. **34**(22): p. 5465-75.

10. Goldberg, R., et al., *Interactions between adsorbed hydrogenated soy phosphatidylcholine (HSPC) vesicles at physiologically high pressures and salt concentrations.* Biophysical Journal, 2011. **100**(10): p. 2403-2411.

11. Klein, J., *Hydration lubrication.* Friction, 2013. **1**(1): p. 1-23.

12. Drew McCormack, S.L.C., Derek Y.C.Chan, *Calculations of Electric Double-Layer Force and Interaction Free Energy between Dissimilar Surfaces.* Journal of Colloid and Interface Science, 1995. **169**(1): p. 177-196.

13. Dimova, R. and K.A. Riske, *Electrodeformation, Electroporation, and Electrofusion of Giant Unilamellar Vesicles*, in *Handbook of Electroporation*. 2016. p. 1-18.

14. Roesel, D., et al., *Ion-Induced Transient Potential Fluctuations Facilitate Pore Formation and Cation Transport through Lipid Membranes.* J Am Chem Soc, 2022. **144**(51): p. 23352-23357.

15. Jin, D. and J. Klein, *Tuning interfacial energy dissipation via topologically electro-convoluted lipid-membrane boundary layers.* Preprint at arXiv:2303.08555 https://arxiv.org/abs/2303.08555 2023.

16. Jin, D., Y. Zhang, and J. Klein, *Electric-field-induced topological changes in multilamellar and in confined lipid membranes.* Preprint at arXiv:2303.08551 https://arxiv.org/abs/2303.08551, 2023.

17. Zhang, Y., et al., *Cell-inspired, massive electromodulation of interfacial energy dissipation.* ArXiv, 2023: p. arXiv:2305.19178v2 [cond-mat.mtrl-sci]

18. Cao, Y., et al., *Normal and shear forces between boundary sphingomyelin layers under aqueous conditions.* Soft Matter, 2020. **16**(16): p. 3973-3980.

19. M'Baye, G., et al., *Liquid ordered and gel phases of lipid bilayers: fluorescent probes reveal close fluidity but different hydration.* Biophys J, 2008. **95**(3): p. 1217-1225.

20. Perkin, S., et al., *Forces between Mica Surfaces, Prepared in Different Ways, Across Aqueous and Nonaqueous Liquids Confined to Molecularly Thin Films.* Langmuir, 2006. **22**(14): p. 6142-6152.

21. Liraz Chai, J.K., *Large Area, Molecularly Smooth (0.2nm rms) Gold Films for Surface Forces and Other Studies.* Langmuir, 2007. **23**(14): p. 7777-7783.

22. Tivony, R., N. Iuster, and J. Klein, *Probing the Surface Properties of Gold at Low Electrolyte*

*Concentration.* Langmuir, 2016. **32**(29): p. 7346-7355.
23. Tivony, R., et al., *Direct Observation of Confinement-Induced Charge Inversion at a Metal Surface.* Langmuir, 2015. **31**(47): p. 12845-9.
24. Israelachvili, J.N., *Thin film studies using multiple-beam interferometry.* Journal of Colloid and Interface Science, 1973. **44**(2): p. 259-272.
25. Clarkson, M.T., *Multiple-beam interferometry with thin metal films and unsymmetrical systems.* Journal of Physics D: Applied Physics, 1989. **22**(4): p. 475-482.
26. Johnson, K.L., et al., *Surface energy and the contact of elastic solids.* Proceedings of the Royal Society of London. A. Mathematical and Physical Sciences, 1971. **324**(1558): p. 301-313.
27. Lin, W. and J. Klein, *Direct measurement of surface forces: Recent advances and insights.* Applied Physics Reviews, 2021. **8**(3): p. 031316.
28. Israelachvili, J.N., *Intermolecular and Surface Forces*. 2011, UK: Elsevier.
29. Kampf, N., et al., *Direct Measurement of Sub-Debye-Length Attraction between Oppositely Charged Surfaces.* Physical Review Letters, 2009. **103**(11): p. 118304.
30. Cao, Y., N. Kampf, and J. Klein, *Boundary Lubrication, Hemifusion, and Self-Healing of Binary Saturated and Monounsaturated Phosphatidylcholine Mixtures.* Langmuir, 2019. **35**(48): p. 15459-15468.
31. Goldberg, R., et al., *Boundary lubricants with exceptionally low friction coefficients based on 2D close-packed phosphatidylcholine liposomes.* Adv Mater, 2011. **23**(31): p. 3517-3521.
32. Lipowsky, R. and U. Seifert, *Adhesion of Vesicles and Membranes.* Molecular Crystals and Liquid Crystals, 2006. **202**(1): p. 17-25.
33. Seifert, U. and R. Lipowsky, *Adhesion of vesicles.* Physical Review A, 1990. **42**(8): p. 4768-4771.
34. Franks, N.P. and W.R. Lieb, *The structure of lipid bilayers and the effects of general anaesthetics: An X-ray and neutron diffraction study.* Journal of Molecular Biology, 1979. **133**(4): p. 469-500.
35. Beerlink, A., et al., *X-ray Structure Analysis of Free-Standing Lipid Membranes Facilitated by Micromachined Apertures.* Langmuir, 2008. **24**(9): p. 4952-4958.
36. Jin, D., et al., *Direct measurement of the viscoelectric effect in water.* Proc Natl Acad Sci U S A, 2022. **119**(1).
37. Deneke, N., A.L. Chau, and C.S. Davis, *Pressure tunable adhesion of rough elastomers.* Soft Matter, 2021. **17**(4): p. 863-869.
38. Sowa, D., Z. Czech, and Ł. Byczyński, *Peel adhesion of acrylic pressure-sensitive adhesives on selected substrates versus their surface energies.* International Journal of Adhesion and Adhesives, 2014. **49**: p. 38-43.
39. Li, L., et al., *Surface Energy and Adhesion Studies on Acrylic Pressure Sensitive Adhesives.* The Journal of Adhesion, 2001. **76**(4): p. 307-334.
40. Sun, Y., C.-C. Lee, and H.W. Huang, *Adhesion and Merging of Lipid Bilayers: A Method for Measuring the Free Energy of Adhesion and Hemifusion.* Biophysical Journal, 2011. **100**(4): p. 987-995.
41. Blokhuis, E.M., et al., *Fusion Pores Live on the Edge.* J Phys Chem Lett, 2020. **11**(4): p. 1204-1208.
42. Frostad, J.M., et al., *Direct measurement of interaction forces between charged multilamellar vesiclesdagger.* Soft Matter, 2014. **10**(39): p. 7769-80.
43. Enoki, T.A. and G.W. Feigenson, *Asymmetric Bilayers by Hemifusion: Method and Leaflet Behaviors.* Biophys J, 2019. **117**(6): p. 1037-1050.

44. Poojari, C.S., K.C. Scherer, and J.S. Hub, *Free energies of membrane stalk formation from a lipidomics perspective.* Nat Commun, 2021. **12**(1): p. 6594.
45. Sorkin, R., et al., *Hydration lubrication and shear-induced self-healing of lipid bilayer boundary lubricants in phosphatidylcholine dispersions.* Soft Matter, 2016. **12**(10): p. 2773-2784.